\documentclass[epj,final]{svjour}

\usepackage{latexsym}
\usepackage{url}
\usepackage{amsfonts}
\usepackage{amsmath, amssymb}
\RequirePackage{graphicx, subcaption}

\usepackage{latexsym}
\usepackage{url}
\usepackage{amsfonts}
\usepackage{amsmath, amssymb}
\RequirePackage{graphicx}
\begin{document}
\title{On the role of $\Lambda$ on accretion disks}
\author{Sh.~Khlghatyan
}                     
%
%
\institute{Center for Cosmology and Astrophysics, Alikhanian National Laboratory and Yerevan State University, Yerevan, Armenia}
\date{Received: date / Revised version: date}
%

\abstract{The dynamics of accretion disk is considered taking into account the Lense-Thirring precession in the presence of cosmological constant $\Lambda$ of Schwarzschild -- de Sitter metric. The nodal and apsidal frequencies are obtained and the role of $\Lambda$ is revealed in their properties, including the consequences for the Bardeen-Petterson effect.
\PACS{
	{PACS-04.50.Kd}{Modified theories of gravity }   
	} 
} 
\maketitle
\section{Introduction}

The accretion disks are known to be crucial structures which exist both in stellar scale in the vicinity of individual stars, and especially, in centers of galaxies, essentially determining the observational properties of those objects \cite{LB,SS,LBP,GO,Bal}. Among the effects influencing the structure and evolution of accretion disks, General Relativity (GR) predictions i.e. Lense-Thirring, Bardeen-Petterson effects \cite{BP} have been considered both for stellar and galactic nuclei scale accretion processes, see \cite{1} and references therein.

At the same time, the observational indications for the dark matter and dark energy have triggered the study of variety of modified gravity models. Among the considered modified gravity models is the one involving the cosmological constant in the weak-field limit of GR  \cite{G,GS1,GS2}. $\Lambda$ acts as second gravity constant along with the Newtonian constant $G$, as one of the fundamental constants of Universe \cite{GS3}. Then the resulting metric is that of Schwarzschild -- de Sitter \cite{R,GS1}
\begin{equation}\label{metric}
g_{tt} = 1 - \frac{2GM}{c^2r} - \frac{\Lambda r^2}{3}, \quad g_{rr} = \left(1 - \frac{2GM}{c^2r} - \frac{\Lambda r^2}{3}\right)^{-1}.
\end{equation}
For the current the numerical value $\Lambda = 1.11 \times 10^{-52} m^{-2}$ \cite{Pl} it is predicted that with an increase in accuracy and statistics in observational data, the observations of gravitational lensing, groups and clusters of galaxies, the flow of galaxies in the vicinity of the Local Group, properties of cosmic voids \cite{GK} can provide informative tests to detect the effects associated with the cosmological constant \cite{GS6}.
The influence of $\Lambda$ on various effects and the possibility to obtain constraints associated with $\Lambda$-term is considered in \cite{L,Con1,GKS,SK,GEM}. Particularly, it was shown that the $\Lambda$-term is able to provide a solution for the Hubble tension problem \cite{GS4}. 

In this paper we study the role of the metric Eq.(1) and of the $\Lambda$-term in the structure and evolution of accretion disks, particularly,  regarding the Lense-Thirring precession and its consequences. We obtain expressions for both nodal and apsidal frequencies with the $\Lambda$ term and reveal the consequences for the Bardeen-Petterson effect \cite{BP}.

\section{Basic equations}

The main equations describing the structure of accretion disks i.e. of mass, angular momentum and energy balance are  \cite{SS,4}
\begin{eqnarray*}
& &\frac{\partial}{\partial t}(2\pi r \Sigma) + \frac{\partial \dot{M}}{\partial r} = 0,\\
& &\frac{\partial}{\partial t}(2\pi r \Sigma j) + \frac{\partial \dot{J}}{\partial r} = 0,\\
& &\frac{\partial}{\partial t}(2\pi r \Sigma e) + \frac{\partial \dot{E}}{\partial r} + 4\pi r F = 0,
\end{eqnarray*}
where $\Sigma$ is the surface mass density, $\dot{M}$ is the rate of accretion, $j$ is the angular momentum per unit mass, $\dot{J}$ is the rate of angular momentum flow, $e$ is the total mechanical energy per unit mass, $\dot{E}$ is the rate at which energy flows across a cylinder with a radius $r$ and $F$ is the rate of energy which is radiated locally from the disk surface.
These quantities are related to the model-dependent parameters of the disk and to the following basic quantities as follows
\begin{equation}
v = \sqrt{r\frac{d\Phi}{dr}},\quad \Omega = \sqrt{\frac{1}{r}\frac{d\Phi}{dr}},\quad j  = \sqrt{r^3\frac{d\Phi}{dr}},
\end{equation} 
where $v$ is the rotational velocity, $\Omega$ is the angular velocity and $\Phi$ is the gravitational potential. Moreover, in order to consider the relativistic effects it is convenient to use the so-called GR potential which is defined as \cite{4}
\begin{equation}\label{PhiGR}
\Phi_{GR} = -\frac{GM}{r-r_g}
\end{equation}
where $r_g=\frac{2GM}{c^2}$ is the gravitational radius.

\section{$\Lambda$ corrections}\label{corr}

For the metric Eq.(\ref{metric}) one can see that $\Lambda$ enters in the weak field limit of GR and at the same time influences in the relativistic precession (or Einstein's precession) \cite{Br}  and the Lense-Thirring effect. In the weak-field limit the gravitational potential is written as 
\begin{equation}\label{Phi}
\Phi = -\frac{GM}{r} - \frac{\Lambda c^2r^2}{6}.
\end{equation} 
On the other hand, for relativistic precession we will get
\begin{equation}\label{OmGR}
\Omega_{GR} = \frac{2GJ}{c^2\tilde{a}^3(1-\tilde{e}^2)^{3/2}}+\frac{\Lambda J}{3M},
\end{equation}
where $\tilde{a}$ and $\tilde{e}$ are the semi-major axis and the eccentricity of the orbit respectively. Then, the Lense-Thirring precession changes to
\begin{equation}\label{Om}
\Omega_{LT} = \frac{2GJ}{c^2r^3}+\frac{\Lambda J}{3M},
\end{equation}
where $J$ is the angular momentum of the central body with mass $M$. 

According to the Eqs.(\ref{Phi}-\ref{Om}) the contribution of the $\Lambda$-term increases at large distances from the object. Namely, the distance beyond which the $\Lambda$-term  becomes dominant is  
\begin{equation}
r^3_{cr} = \frac{6GM}{\Lambda c^2}.
\end{equation}
This critical distance $r_{cr}$ is tabulated in Table \ref{tab1} for certain typical objects. 
\begin{table}[h!]
\caption{}\label{tab1}
\centering
\begin{tabular}{ |p{2.4cm}||p{2.7cm}|p{1.8cm}| }
\hline
\multicolumn{3}{|c|}{Critical distance for different objects} \\
\hline
Central Object& Mass (Kg)&Radius (m)\\
\hline
Earth &5.97 $\times 10^{24}$ & 4.92 $\times 10^{16}$  \\
\hline
Sun & 1.98 $\times 10^{30}=M_{\odot}$ & 4.30 $\times 10^{18}$ \\
\hline
M87${}^{*}$  &6.35 $\times 10^9 M_{\odot}$& 4.06 $\times 10^{21}$ \\
\hline
\end{tabular}
\end{table}

Considering Eq.(\ref{Phi}) and the form of the GR potential in Eq.(\ref{PhiGR}), the main parameters of accretion disk can be written as
\begin{eqnarray}
&&v = \frac{1}{\sqrt{3}}\sqrt{\frac{3GM}{r}\left( 1 + \frac{2r_g}{r}\right)-\Lambda c^2r^2}\\
&&\Omega = \sqrt{\frac{GM}{r}\left(1 +\frac{2r_g}{r}\right)-\frac{\Lambda c^2}{3}}\\
&&j = \sqrt{GMr\left(1 + \frac{2r_g}{r}\right)-\frac{\Lambda c^2r^4}{3}}.\\
\end{eqnarray}

\section{Nodal and apsidal frequencies}

Using the post-Newtonian approximation and taking into account the corrections for the momentum, the equation for the disk takes the following form \cite{1}
\begin{equation}\label{mot}
\frac{d\mathbf{v}}{dt}=-\frac{1}{\rho}\nabla P + \mathbf{v}\times\mathbf{h}-\nabla\Phi +\mathbf{S}_{visc},
\end{equation}
where $S_{visc}$ is the viscous force per unit mass, $\rho$ is the density, $P$ is the pressure and $\mathbf{v}\times\mathbf{h}$ is the gravitomagnetic force according to gravito-electromagnetic (GEM) formalism \cite{GEM1,GEM}. In order to consider the relativistic effects the GR potential in Eq.(\ref{PhiGR}) is used in the above equation for the gravitational potential $\Phi$.

The Lense-Thirring precession causes the plane of an orbit inclined to the $(x, y)$ plane to precess around the angular momentum $(z)$ axis. This precession is taken into account by adding the second term in Eq.(\ref{mot}).
	
The nodal precession frequency that arises due to the inclusion of the gravitomagnetic force term in the equation of motion is defined by the derivative of the effective potential \cite{1,2}
\begin{equation}
\Omega^2 = \frac{\partial^2\Phi_{eff}}{\partial z^2}\Big|_{z=0},
\end{equation}
where the effective potential includes the contribution of the gravitomagnetic force and is given as \cite{3}
\begin{equation}
\Phi_{eff} = \Phi(r) + \Phi',\quad \mathbf{v}\times\mathbf{h}=-\nabla \Phi'.
\end{equation}
	
\section{$\Lambda$-modification}

Considering the corrections given in the section 3, we can obtain the modified nodal and apsidal frequencies. In our case, we are interested in the second and third terms in the Eq.(\ref{mot}), which are modified due to the presence of the $\Lambda$ term. 

Then, $\mathbf{h}$ will take the following form
\begin{equation}\label{hlambda}
\mathbf{h}=\left(\frac{2}{r^3}+\frac{\Lambda}{3M}\right)\mathbf{S}-\frac{6(\mathbf{S}\mathbf{r})}{r^5}\mathbf{r}
\end{equation}
and the gravitational potential is modified as
\begin{equation}
\Phi = -\frac{GM}{r}\left(1 + \frac{r_g}{r}\right) - \frac{\Lambda c^2r^2}{6}.
\end{equation}
Taking into account the modified expressions we find the effective potential 
\begin{equation}\label{eff}
\begin{split}
\Phi_{eff} = &\frac{4S}{5r^{5/2}}\sqrt{GM}\,\, {}_{2}F_{1}\left(-\frac{5}{6},-\frac{1}{2};\frac{1}{6};\frac{\Lambda c^2}{3GM}r^3\right) - \frac{2r}{3}\sqrt{\frac{G}{M}}S\Lambda\,\, {}_{2}F_{1}\left(-\frac{1}{2},\frac{1}{6};\frac{7}{6};\frac{\Lambda c^2}{3GM}r^3\right)\\ &-\frac{3Sz^2}{r^4}\sqrt{\frac{GM}{r}-\frac{\Lambda c^2r^2}{3}}-\frac{GM}{r}\left(1 + \frac{r_g}{r}\right) + \frac{\Lambda c^2r^2}{6},
\end{split}
\end{equation}
where $_{2}F_{1}$ is the hypergeometric function. 

Note, that the cosmological constant $\Lambda$ enters in two places and modifies two phenomena. First, it enters in the potential of the gravitational field $\Phi$ and second, it influences the pure relativistic Lense-Thirring precession frequency. Furthermore, as can be seen from the Eq.(\ref{hlambda}), the additional frequency with the $\Lambda$ term does not depend on the distance or other parameters of the disk.

When $\Lambda \to 0$ the Eq.(\ref{eff}) is reduced to the ordinary Newtonian potential. The difference between the potentials $\Delta \Phi$ due to the presence of $\Lambda$-terms for a black hole of M87* type is shown in Figure 1. One can see that while this difference is negligible for galactic nuclei scales, the presence of $\Lambda$-terms becomes considerable at larger distances. 
\begin{figure}[h!]
\centering
\includegraphics[width=0.6\textwidth]{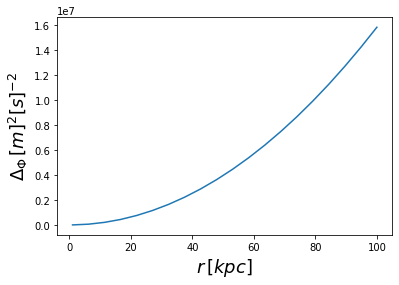}
\caption{The difference between effective potentials for  M87*.}
\end{figure}  

Using the expression for the effective potential and the equation for the nodal frequency $\Omega$ we find the $\Lambda$-modified expression for the nodal precession. In addition, in order to obtain a dimensionless frequency, we normalize the so-called post-Keplerian frequency by taking the $\Lambda$ term into account
\begin{equation}\label{nodal}
\Omega^2 = -\frac{6S}{r^4\omega_{p-K,\Lambda}^{2}}\sqrt{\frac{GM}{r}-\frac{\Lambda c^2r^2}{3}}.
\end{equation}
The post-Keplerian frequency is obtained from the following equation
\begin{equation}
g_{tt,r}\left(\frac{dt}{d\phi}\right)^{2} + 2g_{t\phi,r}\left(\frac{dt}{d\phi}\right) + g_{\phi\phi,r} = 0,
\end{equation}
where $g_{tt}, g_{t\phi}, g_{\phi\phi}$ are the components of the Kerr-de Sitter metric. Accordingly, considering the definition $\omega_{p-K,\Lambda} = \frac{d\phi}{dt}$ we have
\begin{equation}\label{PostKep}
\omega_{p-K,\Lambda}^{2} = \frac{MG}{r^3} - \frac{\Lambda c^2}{3} -\frac{4a}{c}\left(\frac{MG}{r^3} - \frac{\Lambda c^2}{3}\right)^{3/2}.
\end{equation}

Finally, using the nodal frequency according to the Eq.(\ref{nodal}) we find the apsidal frequency
\begin{equation}
\kappa^2 = 4\Omega^2 - \frac{9}{r^5}\frac{\frac{2\Lambda c^2r^3}{3}-3GM}{\sqrt{\frac{GM}{r}-\frac{\Lambda c^2r^2}{3}}}.
\end{equation}

\section{Bardeen-Petterson effect}

The influence of the Lense-Thirring precession on the dynamics of the accretion disk leads to Bardeen-Petterson effect when  the nodal angle changes and as a result, a spiral-like patterns have to appear in the disk \cite{BP}. The rotation angle of the line of nodes is given by
\begin{equation}
\alpha = \frac{r\Omega_{LT}}{v_{r}},
\end{equation}
where $v_r$ is the velocity of the accretion disk that flows into the black hole and consequently it spends time $t=r/v_r$ at radius $r$. For the Shakura-Sunyaev accretion disk \cite{SS} the velocity $v_r$ is
\begin{equation}
v_r = \frac{a}{t_d}
\end{equation}
and $t_d$ is the diffusion time defined as 
\begin{equation}
t_d = \frac{a}{\nu}.
\end{equation}
In the above equation $a$ is the semi-major axis and the diffusion coefficient $\nu$ is related to the rotation angle of nodes $\alpha$, the Keplerian frequency $\omega_K$ and to the scale height $H$ as
\begin{equation}
\nu = \alpha\omega_k H^2.
\end{equation}

Thus we can state that the presence of the $\Lambda$ contributes to the values $\Omega_{LT}$ by the Eq.(\ref{Om}), $\omega_{p-K,\Lambda}$ by Eq.(\ref{PostKep}) and to the scale height $H$. Taking into account all these corrections, we get the following expression for 
\begin{equation}
\alpha^2  = \Omega_{LT}\frac{ar}{\omega_{p-K,\Lambda}}\left(\frac{\mu}{kT}\right)^2\left(\frac{GM}{a^2}-\frac{\Lambda c^2}{3}a\right)^2, 
\end{equation}
where $k$  is the Boltzmann constant, $\mu$ is the mass of the gas particles and the temperature $T$ is obtained by the condition of equilibrium between the heating rate due to viscosity.

\section{Conclusion}

We studied the dynamics of the accretion disk to reveal the contribution of the $\Lambda$-term in the equations of modified GR. We showed that the $\Lambda$-term enters in both relativistic and non-relativistic equations and contributes in two different phenomena accordingly. For the weak-field limit $\Lambda$ enters in the gravitational potential $\Phi$, while for relativistic effects one has the modification due to $\Lambda$ both in Einstein and Lense-Thirring precessions. We obtained the critical distance at which the contribution of $\Lambda$ becomes dominant and by considering the presence of $\Lambda$ in the above mentioned effects we calculated the nodal and apsidal frequencies by solving the relevant equations analytically. For typical astrophysical objects including for accretion disks in galactic centers the real contribution of the $\Lambda$ is negligible. This issue, however, can have sense for consideration in the context of galactic nuclei and galactic larger structure evolutionary interrelations.  One can recall that, initially the gravitational lensing was also considered as non-observable effect.  Finally, we have analyzed the Bardeen-Petterson effect and have shown that due to the presence of the $\Lambda$ in $\Omega_{LT}$ the post-Keplerian frequency $\omega_{p-K,\Lambda}$ and the key parameter, i.e. the rotation angle of the line of nodes are modified.


\begin{thebibliography}{99}

\bibitem{LB} D. Lynden-Bell, Nature \textbf{223}, (1969) 690 
\bibitem{SS} N. I. Shakura, R. A. Sunyaev, A\&A \textbf{24}, (1973) 337
\bibitem{LBP} D. Lynden-Bell, J.E. Pringle, MNRAS \textbf{168}, (1974) 603
\bibitem{GO} V.G. Gurzadyan,  L.M. Ozernoy, A\&A \textbf{95}, (1981) 39
\bibitem{Bal} S.A. Balbus,   Ann. Rev. Astron. Astrophys. \textbf{41}, (2003) 555
\bibitem{BP} J. M. Bardeen, J. A.  Petterson, A\&A \textbf{195}, (1975) L65
\bibitem{1} S. Dyda, C.S. Reynolds, arXiv:2008.12381 (2020)
\bibitem{G} V.G. Gurzadyan,  Eur. Phys. J. Plus \textbf{134}, (2019) 14 
\bibitem{GS1} V.G. Gurzadyan, A. Stepanian,  Eur. Phys. J. C \textbf{78}, (2018) 632
\bibitem{GS2} V.G. Gurzadyan, A. Stepanian,  Eur. Phys. J. C \textbf{79}, (2019) 169
\bibitem{GS3} V.G. Gurzadyan, A. Stepanian, Eur. Phys. J. Plus \textbf{134}, (2019) 98
\bibitem{R} W. Rindler, \textit{Relativity, Special, General, and Cosmological}, (Oxford University Press, 2006)
\bibitem{Pl} P.A.R. Ade et al, A\& A \textbf{594}, (2016) A13
\bibitem{GK} V.G. Gurzadyan, A. Kocharyan, A\&A \textbf{493}, (2009) L61
\bibitem{GS6} V.G. Gurzadyan, A. Stepanian,   Eur. Phys. J. C \textbf{78}, (2018) 869
\bibitem{L} V. Kagramanova, J. Kunz, C. Lammerzahl, Phys. Lett., B \textbf{634}, (2006) 465
\bibitem{Con1}  A.W. Kerr, J.C. Hauck, B. Mashhoon, Class. Quant. Grav. \textbf{20}, (2003) 2727
\bibitem{GKS} V.G. Gurzadyan, A.A. Kocharyan, A. Stepanian, Eur. Phys. J. C \textbf{80}, (2020) 24
\bibitem{SK}  A. Stepanian, Sh. Khlghatyan, Eur. Phys. J. Plus \textbf{135}, (2020) 712
\bibitem{GEM} A. Stepanian, Sh. Khlghatyan, V. G. Gurzadyan, Eur. Phys. J. C \textbf{80}, (2020) 1011
\bibitem{GS4} V.G. Gurzadyan, A. Stepanian,  Eur. Phys. J. C \textbf{79}, (2019) 568
\bibitem{4} B. Paczynsky, P.J. Wiita, A\&A \textbf{88}, (1980) 23
\bibitem{Br} V.A. Brumberg, \textit{Relativistic Celestial Mechanics} (Nauka, Moscow 1972) (in Russian)
\bibitem{GEM1} B. Mashhoon, \textit{Gravitomagnetism}, Proc. XXIII Spanish Relativity Meeting, p. 121, (World Scientific, 2001)
\bibitem{2} R. Nealon, D.J. Price, C.J. Nixon, MNRAS \textbf{448}, (2015) 1526
\bibitem{3} R.P. Nelson, J. Papaloizou, MNRAS \textbf{315}, (2000) 570

\end{thebibliography}
\end{document}